**Gear-tuning meta-shaft for low-frequency torsional vibration suppression**

Dongxian Wang[a], Hao Zhou[a], Jianlei Zhao[a], Zhou Hu[a], Yangyang Chen[c,*], Rui Zhu[a,b,*]

[a] School of Aerospace Engineering, Beijing Institute of Technology, Beijing 100081, China

[b] State Key Laboratory of Environment Characteristics and Effects for Near-space, Beijing Institute of Technology, Beijing 100081, China

[c] Department of Mechanical and Aerospace Engineering, The Hong Kong University of Science and Technology, Clear Water Bay, Hong Kong, China

* Corresponding author

Email: maeychen@ust.hk (Y . Chen), ruizhu@bit.edu.cn (R . Zhu)

# Abstract

Metastructures can be engineered with low-frequency torsional band gaps, which provides a new solution for vibration suppression in shaft systems. However, achieving precise, reversible and robust tunability remains challenging, particularly in shaft systems due to the limited space preventing the implementation of tunable designs with complex mechanisms or additional control units. In this study, a tunable meta-shaft with self-locking gear (SLG) resonators is proposed, where the vibration suppression frequency range of the meta-shaft can be adjusted precisely with a simple gear-tuning mechanism without adding resonator's mass. By shifting the SLG teeth to control the deformation of the six curved beams in the SLG resonators, the torsional stiffness and resonant frequency can be effectively modulated, thereby enabling the generation of tunable low-frequency torsional band gaps. The torsional wave attenuation performance of the meta-shaft with periodically attached SLG resonators is evaluated numerically, and a prototype is fabricated to experimentally verifies its robust and tunable vibration suppression capability. Consistent results among theoretical analysis, numerical simulations, and experiments verify the effectiveness and scalability of the proposed tuning strategy for adaptable torsional vibration control in shaft systems.

**Keywords:** Metastructure; Tunability; Torsional vibration; Self-locking gear; Shaft.

## 1. Introduction

Metastructures have gained widespread attention for their excellent wave attenuation performance within a targeted frequency range [1–4], known as band gaps, and have been extensively investigated in various applications such as mechanical wave filtering [5,6], wave guiding [7,8], vibration suppression [9,10], etc. Band gaps typically originate from either Bragg scattering [11] or local resonance[12,13], with the latter enabling much lower frequency attenuation, nearly two orders of magnitude lower than Bragg scattering[14]. Therefore, many metastructure designs based on the local resonance band gap mechanism have been developed [15,16], where the position and range of the band gap can be customized through the design of local resonator [17,18].

Recent studies on metastructures have primarily focused on the attenuation of transverse, longitudinal, and flexural waves in rods [19,20], beams [21,22], plates [23,24], and pipes [25,26], with growing interest in controlling torsional waves in shaft systems. Yu et al. [27] investigated the propagation of torsional waves in a shaft with periodic local resonators consisting of soft rubber and lead, using real and complex band structures to determine band gaps and torsional wave attenuation performance. Subsequently, Song et al. [28] studied the torsional band gap structure of a periodic shaft composed of hard rings and locally resonant rings, achieving broad combined band gaps at low frequencies. To reduce structural weight, Li et al. [29,30] developed a type of a multi-layered shaft with discretized scatters and investigated the emergence of a low-frequency band gap in torsional vibration. Xiao et al. [31,32] established design criteria for shafts with local resonators and derived closed-form formulas describing the torsional band gap edges and center frequencies. Although the above studies have provided theoretical foundations for the formation of torsional band gaps, generating torsional band gaps at low frequencies often requires heavy and bulky resonators. Moreover, constant torsional band gaps with predetermined positions and widths may be unsuitable for complex and variable environments, greatly limiting the application of metastructures in vibration control engineering applications, thereby making it necessary to consider tunability in the design of metastructures with torsional band gaps.

Due to the fixed material combinations and configurations, it is difficult to design metastructures that achieve vibration attenuation in tunable band gaps without altering the established structure. So far, researchers have achieved tunable band gaps by introducing smart materials/structures or mechanical reconfigurability. Smart materials and structures can adjust band gap characteristics by applying external fields, such as temperature [33–35], electric [36–38], magnetic [39–41], pneumatic actuation [42]. Li et al. [43] realized low-frequency band gap adjustability using temperature-controlled shape memory alloy beams. Ma et al. [44] tuned band gaps by modifying the elastic modulus of piezoelectric elements under a transverse electric field. Montgomer et al. [45] achieved stiffness and band gap variation through deformation-mode branching in a magneto-mechanical metastructure. Liu et al. [46] demonstrated real-time, widely tunable band gaps in a pneumatically actuated mechanical metastructure. However, these approaches typically rely on additional physical fields or complex actuation systems, introducing structural complexity, extra energy consumption, and environmental adaptability issues, which may deteriorate the vibration attenuation performance of the metastructures. Under purely mechanical fields, mechanical reconfigurable metastructures allow post-fabrication band gap tuning through structural adjustments. Han et al. [47] proposed a mechanically reconfigurable memory metastructure based on Kresling origami, enabling tunable vibration attenuation within a desired frequency range. Li et al. [48] developed a configurable tensegrity-based metastructure that achieves tunable band gaps through structural phase transition. In addition, several variable stiffness structures such as linked-spring mechanisms [49], cam-roller-spring systems [50], X-shaped structures [51], and origami-inspired designs [52] have been incorporated into tunable resonators to achieve tunable and low-frequency band gaps in metastructures. Although the above metastructures can effectively achieve vibration attenuation in tunable frequency ranges, these assembled tunable resonators require complex assembly processes with numerous mechanical components, including cams, rollers, springs, and rubber elements, which not only complicate the assembly but also introduce the risk of assembly errors.

Inspired by these issues, integrated variable stiffness structures composed of curved beams have been employed to develop structured tunable resonators, the use of which could significantly reduce the number of components and enable easy fabrication through additive manufacturing without requiring complex assembly, thereby providing a novel design idea for achieving vibration attenuation in tunable and low-frequency band gaps. Moreover, the structured tunable resonators achieve mechanical reconfigurability by adjusting their effective stiffness through applied external forces, enabling resonant frequency modulation and the formation of tunable band gaps in the metastructure. Cai et al. [24] proposed a novel compliant resonator for plate vibration attenuation, achieving low effective stiffness and tunable band gaps via pre-compression. Lin et al. [53,54] developed 2D and 3D metamaterials with structured tunable resonators, revealing the mechanism of low-frequency band gaps and showing that directional pre-compression enables control over band gap ranges. Liu et al. [55] introduced a compact 1D metastructure with truncated conical shells, where varying initial pre-displacements allow tuning of the band gap to lower frequencies for various working conditions. Although previous studies have explored metastructures with structurally tunable resonators for translational and bending vibration attenuation, the tuning mechanisms are often complex, lack robustness, and are difficult to implement efficiently for practical applications. Recently, gear structures that are widely used in various mechanical systems have attracted attention for adjusting the static characteristics of metastructures, such as Young's modulus [56], shear modulus [57], stiffness [58], or elastic-plastic deformation [59]. Only a few recent studies have explored dynamic tuning, and these efforts are mostly limited to multi-mode regulation [60]. However, the integration of gear structures with self-locking functionality into metastructures for torsional band gap tuning and vibration suppression has not yet been explored in shaft systems.

To achieve precise, robust, and tunable torsional vibration suppression in shaft systems, this paper introduces a simplified gear-tuning mechanism to design a tunable resonator, which subsequently enables the development of a gear-tuning meta-shaft capable of shifting its operating frequency between high and low frequency ranges. The remainder of this study is organized as follows: Section 2 describes the configuration design of the gear-tuning meta-shaft. In Section 3, a theoretical model is developed to obtain the effective torsional stiffness of the SLG resonator. Section 4 describes theoretical predictions of the torsional band gap of the gear-tuning meta-shaft, and validates the tunable torsional vibration attenuation region through numerical simulations and experiments. Finally, we present our conclusions in Section 5.

## 2. Conceptual design of the gear-tuning meta-shaft

The frequency range of torsional vibration attenuation in shaft systems is primarily determined by the resonance frequency of the attached resonator. Resonators with tunable stiffness allow modulation of this frequency range, enabling the system to adapt well to environmental variations and effectively suppress low-frequency torsional vibrations. In this context, a gear-tuning meta-shaft with self-locking gear (SLG) resonators is proposed for the first time, which is constructed by periodically arranging the SLG resonators (highlighted by the red dashed box) along a uniform shaft (marked in grey), as shown in Fig. 1(a). The core operating principle of this design lies in strong resonator-shaft coupling near the resonance frequencies of the SLG resonators. When torsional waves propagate from the left end of the meta-shaft, this coupling leads to energy localization and partial dissipation, which effectively attenuates torsional waves within a specific frequency range.

The key design of this meta-shaft lies in the tunable SLG resonator, as shown in the left panel of Fig. 1(b). Each SLG resonator is designed as an integrated unit comprising two mirrored SLG sub-resonators, rigidly fixed to the shaft via a central ring (marked in orange) to ensure structural stability. Two sub-resonators address two key requirements: providing low stiffness and enabling precise tunability. This is accomplished through three fundamental components: 1) a set of three identical curved beams (marked in yellow) acting as the primary elements, 2) a circular frame (marked in cyan or green) serving as the resonant mass, and 3) SLG teeth uniformly distributed on one side of the frame. The curved beams are strategically designed as variable-stiffness structures, whose elastic deformation provides the low-stiffness characteristics essential for low-frequency resonance. Additionally, the SLG mechanism, evolved from conventional gear systems widely used in industrial design, enables precise control of angular displacement through tooth engagement and secure connection via a groove-based self-locking mechanism. By rotating the two SLG sub-resonators in opposite directions, the SLG teeth at the edges of the resonators are shifted, activating the self-locking mechanism to prevent relative sliding between the SLG sub-resonators. Color-coded teeth are used in the middle panel of Fig. 1(b) to illustrate the positional changes induced by the relative sliding. The sliding motion forces all six curved beams in each SLG resonator into a state of increased elastic deformation, causing the overall torsional stiffness of the SLG resonator to gradually decrease with increased tooth shifting, as shown in the right panel of Fig. 1(b). This direct control over stiffness allows for precisely adjusting the resonator's resonant frequency post-fabrication, enabling the system to adapt to different operational conditions.

Fig. 1(c) schematically demonstrates the tunable torsional vibration attenuation performance of the proposed meta-shaft by evaluating its transmissibility. The presence of the attenuation region is indicated by the frequency range where the transmissibility falls below 0 dB, signifying that the output vibration signal amplitude is lower than the input excitation. By comparing the transmissibility curves of three states (initial, one tooth shifted, and three teeth shifted), we observe that as the number of tooth shifts increases, the starting frequency of vibration attenuation, progressively shifts toward lower frequencies. This clear leftward shift of the attenuation region confirms that the gear-tuning mechanism effectively lowers the resonant frequency, thereby extending its suppression capability into the low-frequency range. The underlying physical mechanism is visualized in the side views of the gear-tuning meta-shaft, which show how the relative rotation of the SLG sub-resonators deliberately deforms the curved beams, reducing their effective torsional stiffness and enabling precise frequency tuning.

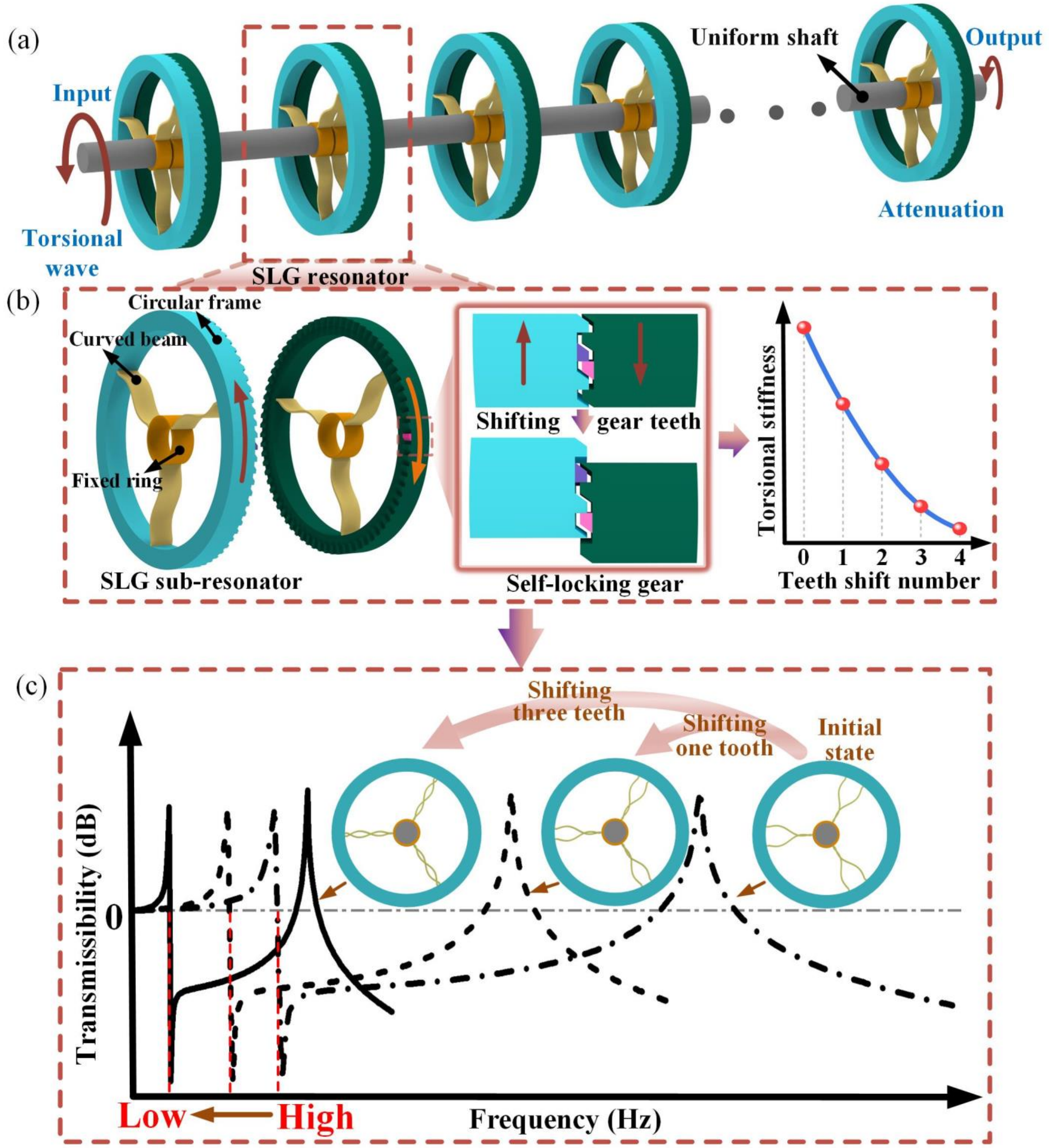

**Fig. 1.** Conceptual design of the gear-tuning meta-shaft. (a) Physical model of the gear-tuning meta-shaft comprising

periodically arranged SLG resonators (highlighted by the red dashed box) and a uniform shaft (marked in grey). (b) Components of the SLG resonator model and the gear-tuning mechanism. Left panel: two mirrored SLG sub-resonators. Middle panel: gear-tuning mechanism. Right panel: relationship between the effective torsional stiffness of the SLG resonator and the teeth shift number. (c) Schematic diagram of transmissibility curve for the gear-tuning meta-shaft. The transmissibility is defined as $20\log_{10}(A_{out}/A_{in})$, where $A_{in}$ and $A_{out}$ represent amplitudes of angular acceleration at the input and output of the meta-shaft, respectively. The transmissibility curves with different line styles correspond to the initial state, one tooth shift, and three teeth shift of the SLG resonators, respectively.

## 3. Static modeling and analysis

The curved beam, as a key component of the SLG resonator, is crucial for coupling externally generated torsional energy to the local resonance within the structure, facilitating energy localization and dissipation. Its inherent geometric nonlinearity allows for a broad range of stiffness tunability, enabling the gear-tuning mechanism to induce substantial variations in the effective torsional stiffness of the SLG resonator. In this section, we focus on the development of an analytical model to evaluate the effective torsional stiffness of the SLG resonator, specifically through the analysis of a curved beam-shaft unit, the fundamental functional unit that directly reflects the overall stiffness behavior of the resonator. An analytical static model of this curved beam-shaft unit is established using the chained beam constraint model (CBCM), which captures the tunable stiffness characteristics. This model is then validated through both numerical simulations and experiments. Finally, the effective torsional stiffness of the SLG resonator is determined, providing the foundation for the following dynamic analysis of the gear-tuning meta-shaft.

### *3.1 Theoretical analysis*

A single SLG sub-resonator together with the attached uniform shaft is shown in the left panel of Fig. 2. Each SLG sub-resonator contains 60 SLG teeth uniformly distributed along one side. Due to the identical geometry of the three curved beams in the SLG sub-resonator, the curved beam-shaft unit with a representative curved beam is designed and analyzed, as shown in the right panel Fig. 2, and it consists of a curved beam, a fixed ring, and a part of uniform shaft. The grey circle represents the cross-section of the uniform shaft with radius $r$ and centered at point A. The orange annulus denotes the cross-section of the fixed ring with thickness $b$, which intersects the curved beam at point Q on its outer circumference. The initial inclination angle $\alpha$ between the centerline of the curved beam and the horizontal axis at point Q is a critical parameter that determines the torsional stiffness characteristics of the curved beam-shaft unit.

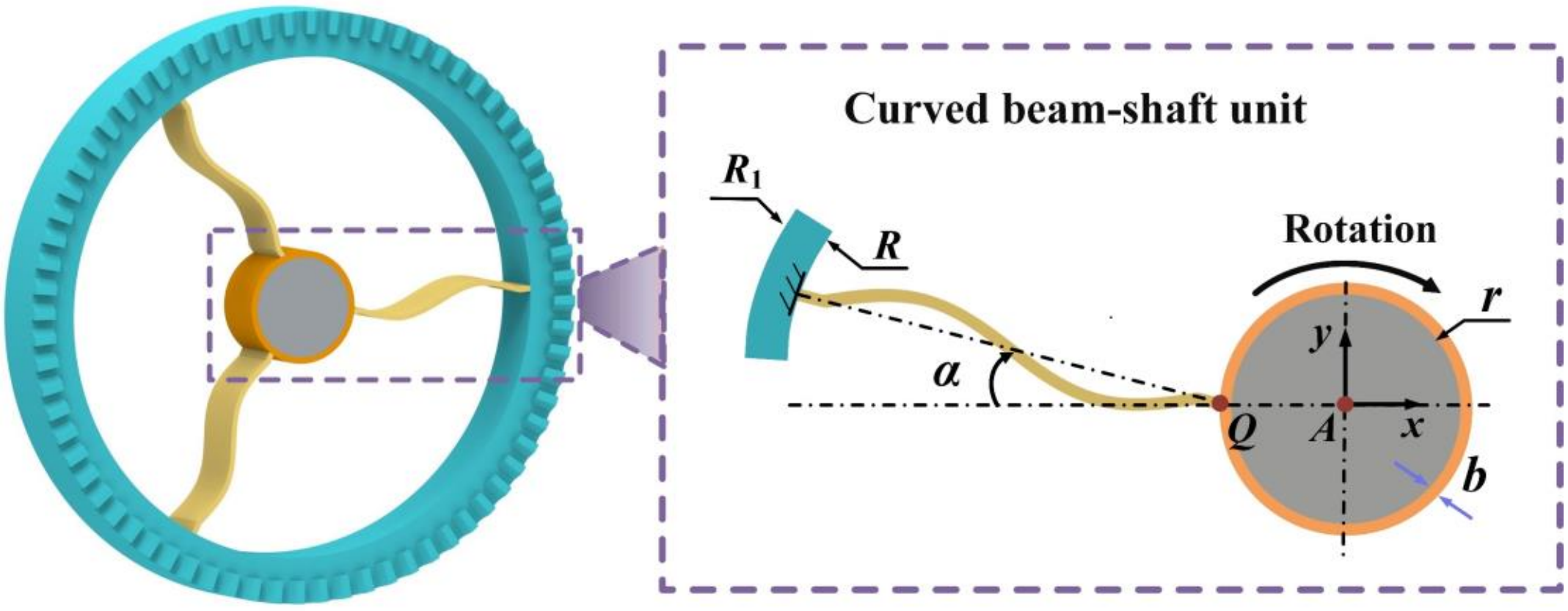

**Fig. 2.** Left panel: schematic of a SLG sub-resonator together with part of the uniform shaft (marked in grey). Right panel: Geometric schematic of the curved beam-shaft unit and its relation to other SLG sub-resonator components.

The initial shape of the curved beam in the curved beam-shaft unit is represented as a piecewise function, which makes analytically determining its torsional stiffness challenging. To address this, we develop an analytical model that characterizes the torsional stiffness by leveraging the capabilities of the CBCM [61–64] in handling irregular structural configurations. The theoretical model, shown in Fig. 3(a), assumes that the curved beam follows a circular path, guided by the fixed ring and uniform shaft. Since the fixed ring and the shaft are rigidly connected to form a relatively thick central structure, they are collectively assumed to act as a rigid beam. The right endpoint of the curved beam is connected to the rigid beam at point Q. As the rigid beam rotates about the center point A, the tip at point Q of the curved beam follows a circular arc path. Fig. 3(b) also illustrates the configuration of the curved beam, whose central part is designed as a sine curved segment, with straight-line sections added at both ends for smooth transition. The parameters $h$, $t$, $l_1$, $l_2$, and $W$ stand for the apex height, in-plane thickness, straight-line length, sine curve length, and out-plane thickness of the curved beam, respectively. For convenience, the initial profile of the curved beam is established in the local coordinate system ($OXY$) as follows:

$$Y(X)=\begin{cases} 0 & X\in[0,l_1] \\ \dfrac{h}{2}\sin(2\pi(X-l_1)/l_2), & X\in(l_1,l_1+l_2] \\ 0 & X\in(l_1+l_2,l] \end{cases} \tag{1}$$

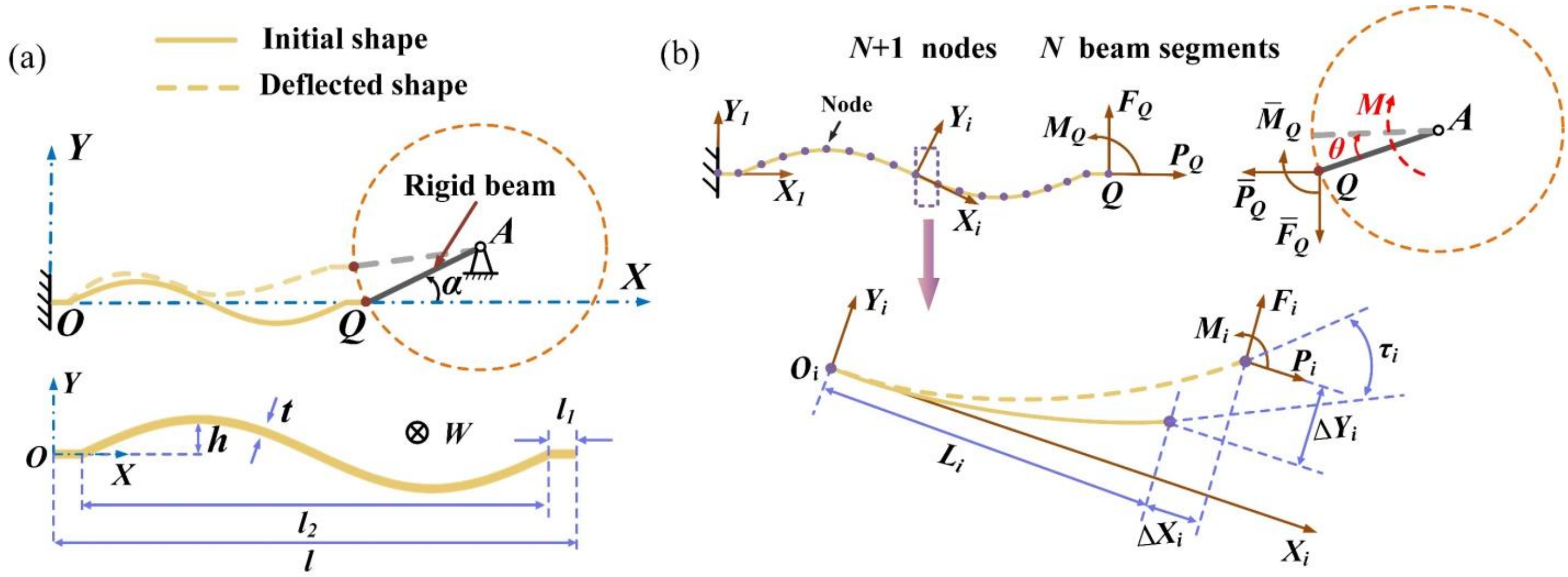

**Fig. 3.** Theoretical model of static. (a) Static model of the curved beam-shaft unit. The model is assumed to be a circular path-guided mechanism. (b) Mechanical analysis of the curved beam and the rigid beam. The curved beam is assumed to be equally divided into $N$ beam segments with $N$+1 nodes (marked as purple dots), and local coordinate system ($O_iX_iY_i$) together with the loading configuration for the $i$-th beam segment.

The core of the CBCM is the discretization of continuous bodies into multiple elements. As shown in Fig. 3(b), the curved beam is assumed to be equally divided into $N$ beam segments with $N$+1 nodes (marked as purple dots), each associated with a local coordinate system ($O_iX_iY_i$). These $N$+1 nodes are represented in terms of their coordinates along the $Y$-axis, slope angle, and curvature as follows:

$$Y_i = Y(X_i) \tag{2}$$

$$\beta_i = \arctan Y'(X_i) \tag{3}$$

$$K_i = \frac{Y''(X_i)}{\left[1+Y'^2(X_i)\right]^{3/2}} \quad (i=1,2,3,\ldots,N+1) \tag{4}$$

The length, normalized thickness, and normalized curvature of each beam segment are given as follows:

$$L_i = (X_{i+1}-X_i)\cos\beta_i + (Y_{i+1}-Y_i)\sin\beta_i \tag{5}$$

$$t_i = \frac{T}{L_i} \tag{6}$$

$$\kappa_i \approx \frac{K_i + K_{i+1}}{2} L_i \quad (i = 1, 2, 3, ..., N) \tag{7}$$

where $T$ represents the uniform thickness of the curved beam.

Furthermore, in the local coordinate system ($O_iX_iY_i$), the local loads $P_i$, $F_i$, and $M_i$ at the right end of $i$-th beam segment represent the forces along the $O_iX_i$ and $O_iY_i$ directions, as well as the moment, as shown in Fig. 3(b). The $\Delta X_i$, $\Delta Y_i$, and $\tau_i$ respectively denote the corresponding deformations, together with the end rotation angle. All these parameters are normalized as follows:

$$p_i = \frac{P_i L_i^2}{EI},\ f_i = \frac{F_i L_i^2}{EI}, m_i = \frac{M_i L_i}{EI}, \delta_i = \frac{\Delta X_i}{L_i}, \lambda_i = \frac{\Delta Y_i}{L_i} \tag{8}$$

where $E$ is the Young's modulus and $I$ is the cross-sectional moment of inertia.

Based on the beam-column theory, the closed-form equations for the beam segments are derived using load-equilibrium and linearized curvature assumptions, and can be expressed as follows [61–63]:

$$\begin{aligned} \begin{bmatrix} f_i \\ m_i \end{bmatrix} &= \begin{bmatrix} g_{11} & g_{12} \\ g_{21} & g_{22} \end{bmatrix} \begin{bmatrix} \lambda_i \\ \tau_i \end{bmatrix} + p_i \begin{bmatrix} p_{11} & p_{12} \\ p_{21} & p_{22} \end{bmatrix} \begin{bmatrix} \lambda_i \\ \tau_i \end{bmatrix} \\ &+ p_i^2 \begin{bmatrix} q_{11} & q_{12} \\ q_{21} & q_{22} \end{bmatrix} \begin{bmatrix} \lambda_i \\ \tau_i \end{bmatrix} + p_i \begin{bmatrix} \kappa_i / 2 \\ \kappa_i / 12 \end{bmatrix} \end{aligned} \tag{9}$$

$$\begin{aligned} \delta_i &= \frac{t_i^2 p_i}{12} - \frac{\kappa_i}{2}\delta_i - \frac{\kappa_i}{12}\tau_i - \frac{1}{2}\begin{bmatrix} \delta_i & \tau_i \end{bmatrix} \begin{bmatrix} p_{11} & p_{12} \\ p_{21} & p_{22} \end{bmatrix} \begin{bmatrix} \delta_i \\ \tau_i \end{bmatrix} \\ &- p_i \begin{bmatrix} \delta_i & \tau_i \end{bmatrix} \begin{bmatrix} q_{11} & q_{12} \\ q_{21} & q_{22} \end{bmatrix} \begin{bmatrix} \delta_i \\ \tau_i \end{bmatrix} + p_i \frac{\kappa_i}{360}\tau_i + p_i \frac{\kappa_i^2}{720} \end{aligned} \tag{10}$$

where the dimensionless coefficients of the CBCM are listed in Table 1.

**Table 1**

Non-dimensional characteristic coefficients of the CBCM matrices.

| $g_{11}$ | $g_{12}$=$g_{21}$ | $g_{22}$ | $p_{11}$ | $p_{12}$=$p_{21}$ | $p_{22}$ | $q_{11}$ | $q_{12}$=$q_{21}$ | $q_{22}$ |
|---|---|---|---|---|---|---|---|---|
| 12 | -6 | 4 | 6/5 | -1/10 | 2/15 | -1/700 | 1/1400 | -11/6300 |

The force equilibrium equations of the curved beam and the first beam segment are given as:

$$P_Q = P_1 = \frac{p_1 EI}{L_1^2},\quad F_Q = F_1 = \frac{f_1 EI}{L_1^2} \tag{11}$$

and since the right end of the $N$th beam segment coincides with the tip of the curved beam, the following relation can be obtained:

$$M_Q = M_N = \frac{m_N EI}{L_N} \tag{12}$$

where $P_Q$, $F_Q$ and $M_Q$ denote three load parameters in the global coordinate system (*OXY*) at the tip of the curved beam.

The force equilibrium equations from the first to the *i*-th beam segment ($i$ = 2,3,…,$N$) are given as follows:

$$\begin{cases} \frac{f_1 EI}{L_1^2}\cos\theta_1 + \frac{p_1 EI}{L_1^2}\sin\theta_1 = \frac{f_i EI}{L_i^2}\cos\theta_i + \frac{p_i EI}{L_i^2}\sin\theta_i \\ -\frac{f_1 EI}{L_1^2}\sin\theta_1 + \frac{p_1 EI}{L_1^2}\cos\theta_1 = -\frac{f_i EI}{L_i^2}\sin\theta_i + \frac{p_i EI}{L_i^2}\cos\theta_i \end{cases} \tag{13}$$

where $\theta_i$ is the rotation angle of the *i*-th beam segment relative to the coordinate system (*OXY*), can be expressed as:

$$\theta_1 = 0, \quad \theta_i = \beta_i + \sum_{k=1}^{i-1}\tau_k \left(i = 2,3,...,N\right) \tag{14}$$

Besides, the moment equilibrium equation for the *i*-th beam segment ($i$ = 2,3,…,$N$) is given by:

$$\frac{m_{i-1}EI}{L_{i-1}} = \frac{m_i EI}{L_i} + \frac{f_i EI}{L_i^2} L_i\left(1+\lambda_i\right) - \frac{p_i EI}{L_i^2} L_i\left(0.5\kappa_i + \delta_i\right) \tag{15}$$

The geometric equations of the curved beam can be expressed as:

$$\begin{cases} \sum_{i=1}^{N}\left[\left(1+\delta_i\right)L_i\cos\theta_i - \left(0.5\kappa_i + \lambda_i\right)L_i\sin\theta_i\right] = X_Q \\ \sum_{i=1}^{N}\left[\left(1+\delta_i\right)L_i\sin\theta_i + \left(0.5\kappa_i + \lambda_i\right)L_i\cos\theta_i\right] = Y_Q \\ \beta_{N+1} + \sum_{i=1}^{N}\tau_i = \theta_Q \end{cases} \tag{16}$$

where $\theta_Q$ is its rotation angle relative to the *X*-axis, $X_Q$ and $Y_Q$ are the coordinates of the curved beam tip.

The rigid beam is connected to the curved beam at point Q, and the coordinates of the curved beam tip can be obtained from the geometric relationship of the rigid beam, as follows:

$$X_Q = l + \left(r+b\right)\cos\left(\alpha\right) - \left(r+b\right)\cos\left(\alpha-\theta\right), \quad Y_Q = \left(r+b\right)\sin\left(\alpha-\theta\right), \quad \theta_Q = -\theta \tag{17}$$

where $l$ denotes the total length of the curved beam, $r$ is the radius of the uniform shaft, $b$ is the thickness of the fixed ring, and ($r$+$b$) represents the length of the rigid beam in the model.

When the curved beam is discretized into $N$ beam segments, 6$N$ unknown variables are introduced, and at least 6$N$ equations must be established to obtain the complete solution. These requirements are fulfilled by Eq. (9)–(16), which constitute the CBCM formulations for the curved beam. Among three load parameters $P_Q$, $F_Q$, and $M_Q$ and three deflection parameters $X_Q$, $Y_Q$, and $\theta_Q$ at the tip of the curved beam, given any set of three parameters, the remaining three parameters can be determined by numerically solving the CBCM formulations.

As the tip loads of the rigid beam and the curved beam at point Q are interaction forces, the tip loads of the rigid beam are defined as $\bar{P}_Q$, $\bar{F}_Q$, and $\bar{M}_Q$, expressed as:

$$\bar{P}_Q = -P_Q, \quad \bar{F}_Q = -F_Q, \quad \bar{M}_Q = -M_Q \tag{18}$$

When the rigid beam rotates about point A, the torque acting on point A is jointly determined by $\bar{P}_Q$, $\bar{F}_Q$, and $\bar{M}_Q$. Accordingly, the relationship between the rotation angle $\theta$ of the rigid beam in the curved beam-shaft unit and its restoring torque $M$ can be expressed as:

$$M = \bar{M}_Q - \bar{F}_Q(r+b)\cos(\alpha-\theta) + \bar{P}_Q(r+b)\sin(\alpha-\theta) \tag{19}$$

The torsional stiffness $K$ of the curved beam-shaft unit, defined as the derivative of the restoring torque $M$ with respect to $\theta$, can be expressed as:

$$K = \frac{\mathrm{d}M}{\mathrm{d}\theta} \tag{20}$$

### *3.2. Results and validation of the static numerical simulations and experiments*

This section introduces numerical simulations and experiments to validate the theoretical model and further characterize the torsional stiffness of the curved beam-shaft unit. First, a 3D Finite Element (FE) model of the curved beam-shaft unit is established using the commercial FE software ANSYS to analyze its nonlinear torsional stiffness characteristics, as shown in Fig. 4. The structural parameters of the curved beam-shaft unit and other components are listed in Table 2. To ensure consistency with the experimental study and consider the additive manufacturing process, photosensitive resin (with a Young's modulus of 2.475 GPa, Poisson's ratio of 0.33, density of 1380 kg/m3) is selected as the material for the curved beam-shaft unit. The 3D solid elements of 8-Node are used in the model, with mesh details enlarged in Fig. 4, and a mesh convergence analysis is performed to ensure the accuracy of the numerical simulation. A finer mesh is applied to the curved beam due to its complex deformation, compared to other components of the model. Specifically, a mesh size of 0.1 mm is assigned along both the length and width of the curved beam, and six meshes are arranged along the thickness direction. Besides, a uniform mesh size of 0.2 mm is used for the fixed ring, with three meshes are arranged along the thickness direction, and the uniform shaft is discretized using a mesh size of 0.5 mm. The analysis is conducted using a static geometrically nonlinear algorithm, and a bonded contact is applied between the fixed ring and the uniform shaft, and the left end of the curved beam is fully fixed. Rotation angles about the negative z-axis are applied to all nodes on the outer surface of the fixed ring to actuate the structure. The restoring torque $M$ of all nodes on the outer surface of the fixed ring is extracted by incorporating the rotation angle $\theta$, allowing the $M$–$\theta$ curves of the curved beam-shaft unit to be obtained.

**Table 2**
Parameters of the SLG resonator.

| Parameters | Descriptions | Values |
|---|---|---|
| $r$ | The radius of the uniform shaft | 5 mm |
| $R$ | The inner radius of the circular frame | 23.6mm |
| $R_1$ | The outer radius of the circular frame | 27.6 mm |
| $l$ | The total length of the curved beam | 16 mm |
| $l_1$ | The connection length of the curved beam | 1 mm |
| $h$ | The apex height of the curved beam | 1.1 mm |
| $t$ | The in-plane thickness of the curved beam | 0.5 mm |
| $\alpha$ | Initial inclination angle | 0.17 rad |
| $b$ | The thickness of the fixed ring | 0.6 mm |
| $W$ | The out-plane thickness of the curved beam | 5 mm |

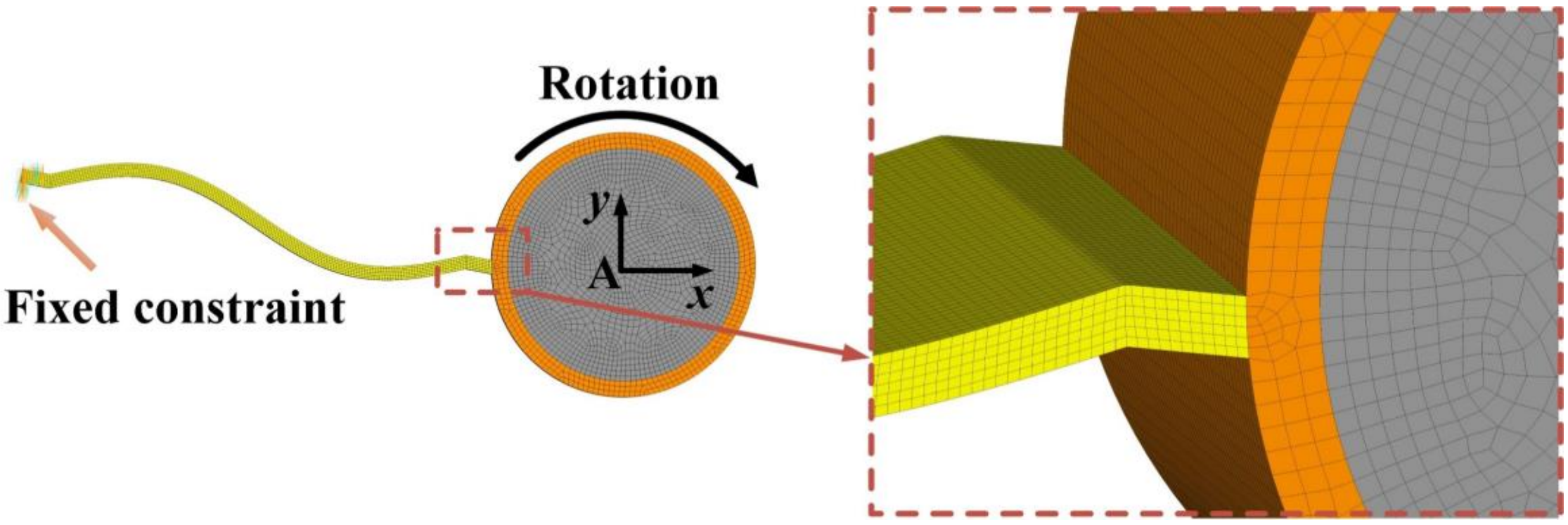

**Fig. 4.** FE model and boundary conditions of the curved beam-shaft unit for static numerical simulations.

Subsequently, a torsional stiffness testing platform was established to validate the stiffness characteristics of the curved beam-shaft unit, with the geometric and material parameters consistent with those described above. Fig. 5(a) shows the instruments and testing system used in the static experiments. To enhance stability during torque measurement and ensure consistency with dynamic analysis, three identical curved beams are arranged symmetrically around the center with a motor interface, minimizing issues related to robustness and assembly eccentricities. The digital torque meter is connected to the connecting shaft, test sample, and motor. We note that the alignment of the central axes among the digital torque meter hole, test sample, connecting shaft, and motor rotation shaft along the same horizontal line is crucial. The thicker end of the connecting shaft is inserted into the hole of the digital torque meter, and the thinner end is inserted into the square hole of the test sample. The accuracy and measurement range of the digital torque meter are $1\times10^{-3}$ N·m and 0 ~ 1 N·m, respectively. The rotation angle excitation required for the static experiments is provided by a constant-speed motor operating at 0.15 rev/min (0.9425 rad/min) under load conditions. The restoring torque generates by the test sample is transmitted to the digital torque meter via the connecting shaft, and a computer program continuously monitors and records the restoring torque and the corresponding time. The low speed of the motor maintained quasi-static conditions, and the measured time was converted into rotation angle based on its constant speed. To enable comparison with the theoretical analysis and simulation results, the measured restoring torque of the test sample was divided by three, yielding the restoring torque of curved beam-shaft unit and thus the corresponding $M$–$\theta$ relationship.

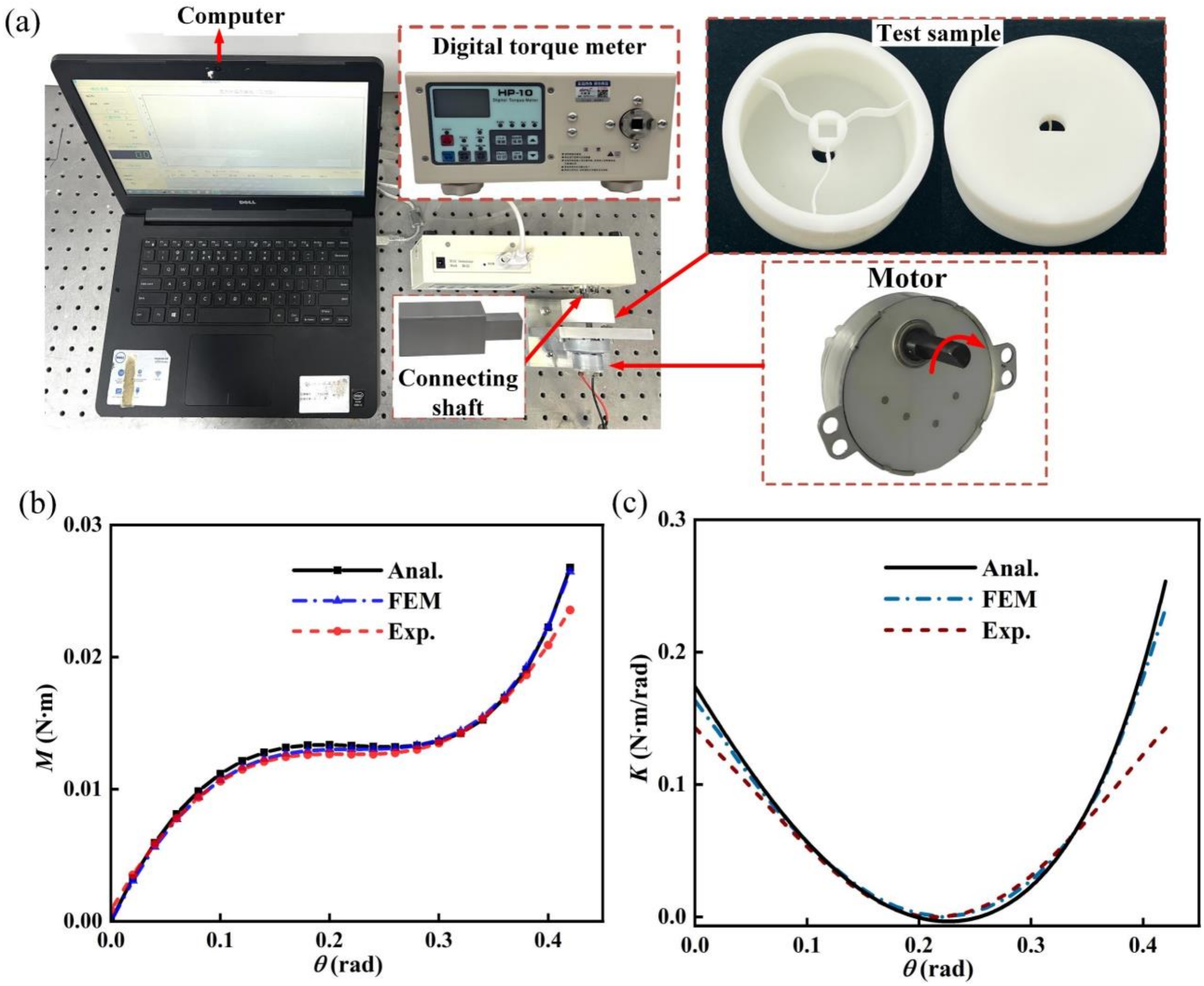

**Fig. 5.** Static experiments and results analysis. (a) Experimental setup for the static experiments, including motor, connecting shaft, test sample, digital torque meter, and computer. Theoretical analysis, FE simulations, and experiments, of (b) the restoring torque *M*–rotation angle $\theta$ curve and (c) torsional stiffness *K*–rotation angle $\theta$ curve.

As shown in Fig. 5(b), the $M$–$\theta$ curves of the curved beam-shaft unit obtained from theoretical analysis, FE simulations, and experiments are compared. Strong consistency is observed among the three results, validating the accuracy of the static model. The restoring torque first increases with the rotation angle, then remains almost unchanged over a certain range, and finally increases again. The nearly constant restoring torque means that the torsional stiffness of the curved beam-shaft unit approaches zero. The corresponding $K$–$\theta$ curves of the curved beam-shaft unit are shown in Fig. 5(c), where the lowest torsional stiffness $5.3\times10^{-4}$ N·m·rad$^{-1}$ occurs at a rotation angle of 0.21 rad. Nevertheless, compared with the simulation and experimental results, the theoretical $M$–$\theta$ curve is slightly higher in the initial stage, mainly due to the simplifying assumption in the static model where the central structure is idealized as a rigid beam. However, the static model still succeeds in capturing the overall trend and torsional stiffness characteristics of the curved beam-shaft unit with sufficient accuracy. In the large-deformation stage, the experimental *M–θ* curve is slightly below the FE simulation prediction, mainly due to differences in boundary conditions, particularly the deformation at the interface between the motor shaft and the resin-based test sample.

### *3.3. Effects of the geometric parameters*

Subsequently, the influence of several fundamental geometric parameters, such as the initial inclination angle $\alpha$, the apex height $h$, and the in-plane thickness $t$ on the torsional stiffness characteristics of the curved beam-shaft unit is determined based on theoretical analysis. During the theoretical analysis, the fundamental geometric parameters in Table 2 remain unchanged, while the initial inclination angle is varied across four different values. The restoring torque $M$ and torsional stiffness $K$ of the curved beam-shaft unit for different initial inclination angles $\alpha$ (0 rad, 0.10 rad, 0.17 rad, and 0.27 rad) are shown in Fig. 6(a). As the initial inclination angle increases, the curved beam-shaft unit sequentially exhibits different torsional stiffness characteristics: positive torsional stiffness, near-zero torsional stiffness, and negative torsional stiffness. Additionally, changing only the apex height $h$ leads to similar conclusions, as shown in Fig. 6(b). The effective torsional stiffness of the SLG resonator is directly related to the torsional stiffness characteristics of the curved beam-shaft unit. If the torsional stiffness at the lowest point of the *K–θ* curve is too high, low-frequency torsional vibration attenuation cannot be achieved. Moreover, the occurrence of negative stiffness would lead to structural instability. Therefore, when the torsional stiffness of the curved beam-shaft unit approaches zero, it becomes more favorable for the SLG resonator to open low-frequency torsional band gaps, thus achieving superior torsional vibration attenuation performance.

As shown in Fig. 6(c), when only changing the in-plane thickness $t$ of the curved beam, the *M–θ* curves improve significantly, indicating that a higher torque is required for the curved beam-shaft unit to tune its torsional stiffness. Furthermore, the *K–θ* curves of the curved beam-shaft unit nearly coincide around the lowest torsional stiffness point. However, the $K$–$\theta$ curves of curved beam-shaft unit become steeper as the in-plane thickness increases, indicating that the region where the torsional stiffness approaches zero gradually decreases as the in-plane thickness increases. The torsional stiffness characteristics of the designed curved beam-shaft unit can be effectively tailored by adjusting the fundamental geometric parameters of the curved beams, allowing rapid alignment with the functional demands of subsequent resonator designs, and thereby broadening its applicability in resonator systems.

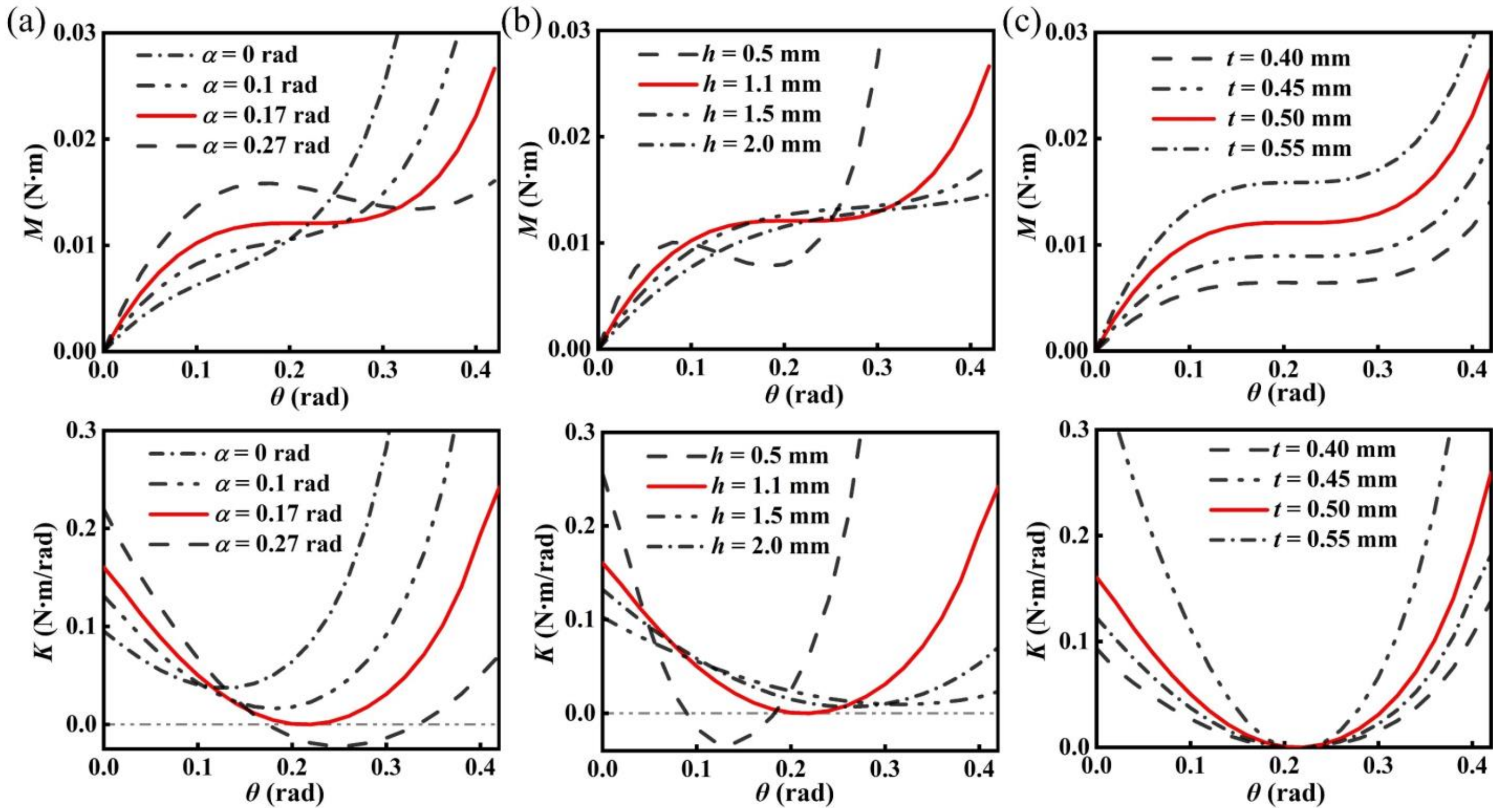

**Fig. 6.** Geometric parameter effects of the curved beam-shaft unit. The restoring torque $M$ and torsional stiffness $K$ for (a) different initial inclination angles $\alpha$ (0 rad, 0.1 rad, 0.17 rad, and 0.27 rad), (b) different apex heights $h$ (0.5 mm, 1.1 mm, 1.5 mm, and 2 mm) of the curved beam, and (c) different in-plane thicknesses $t$ (0.40 mm,0.45 mm, 0.50 mm, and 0.55 mm) of the curved beam.

### *3.4. Effective torsional stiffness of the SLG resonator*

The curved beam-shaft unit is the fundamental functional element of the SLG resonator, and its static model characterizes the variation of torsional stiffness with rotation angle, thereby establishing the direct relationship between the resonator's effective torsional stiffness and the gear-tuning angle. Given that the resonant frequency of the SLG resonator is intrinsically linked to its effective torsional stiffness, accurate determination of this stiffness is of critical importance for predicting and tailoring the dynamic response of the gear-tuning meta-shaft.

The stiffness analysis model of the SLG resonator, which consists of an SLG resonator and the attached uniform shaft, is shown in Fig. 7(a). The entire SLG resonator comprises 60 pairs of SLG teeth, corresponding to a central angle of 0.105 rad (6 degrees) for each pair of SLG teeth. From this, it can be inferred that when the SLG is shifted by one tooth ($\eta$ = 1), the two SLG sub-resonators will rotate by 0.0525 rad (3 degrees) in opposite directions. The figure further illustrates that the SLG teeth can be shifted by applying external forces in opposite rotational directions, while in the absence of external force, the SLG remains in a self-locking state. Moreover, when the SLG is shifted by four teeth ($\eta$ = 4, i.e., the circular frame rotates by 0.21 rad), the SLG resonator will exhibit an extremely small value of torsional stiffness, which has been demonstrated in Fig. 5(c).

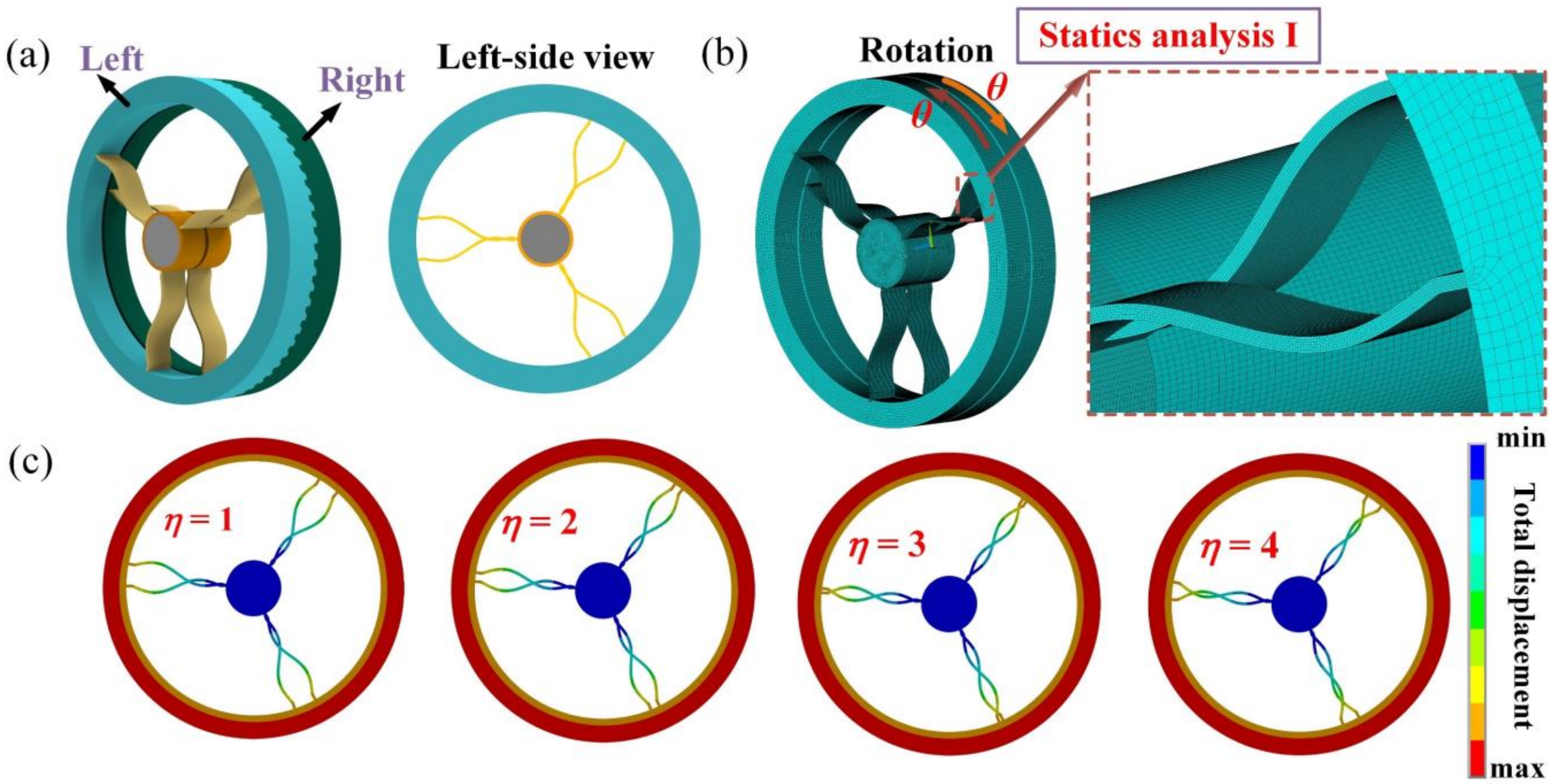

**Fig. 7.** Stiffness analysis model of the SLG resonator and its corresponding results. (a) Schematic diagram of the SLG resonator. Two circular frames are marked in cyan and green for distinction. (b) FE model of the SLG resonator for the first step static analysis. (c) Total displacement field of the SLG resonator with different SLG teeth shifted numbers $\eta$.

To obtain the effective torsional stiffness $K_e$ of the SLG resonator for different SLG teeth shift numbers $\eta$, a two-step static analysis is performed on the SLG resonator. In the first step, a 3D FE model is established using the commercial FE software ANSYS, as shown in Fig. 7(b). Following the convergence analysis and considering computational efficiency, the element types and mesh sizes of the curved beam, fixed ring, and uniform shaft are kept consistent with those in Section 3.2. Meanwhile, rotation angles about the positive $z$-axis are applied to all nodes on the surface of the circular frame (marked in cyan) in the left sub-resonator, while rotation angles about the negative $z$-axis are applied to all nodes on the surface of the circular frame (marked in green) in the right sub-resonator. The deformation fields of the SLG resonator in the side view for different values of $\eta$ are presented in Fig. 7(c). It can be observed that different SLG teeth shift numbers $\eta$ lead to noticeable variations in the deformation states and positional configurations of the curved beams within SLG resonator, thereby reflecting the tunable stiffness characteristics of the SLG resonators and their influence on the overall torsional response of the gear-tuning meta-shaft. Additionally, it is also observed that the two sub-resonators exhibit identical torsional responses but in opposite directions. This symmetry ensures that when the circular frames are rotated in opposite directions, the torques exerted on the uniform shaft are balanced, allowing it to maintain a state of static equilibrium without inducing undesired net rotation or lateral displacement.

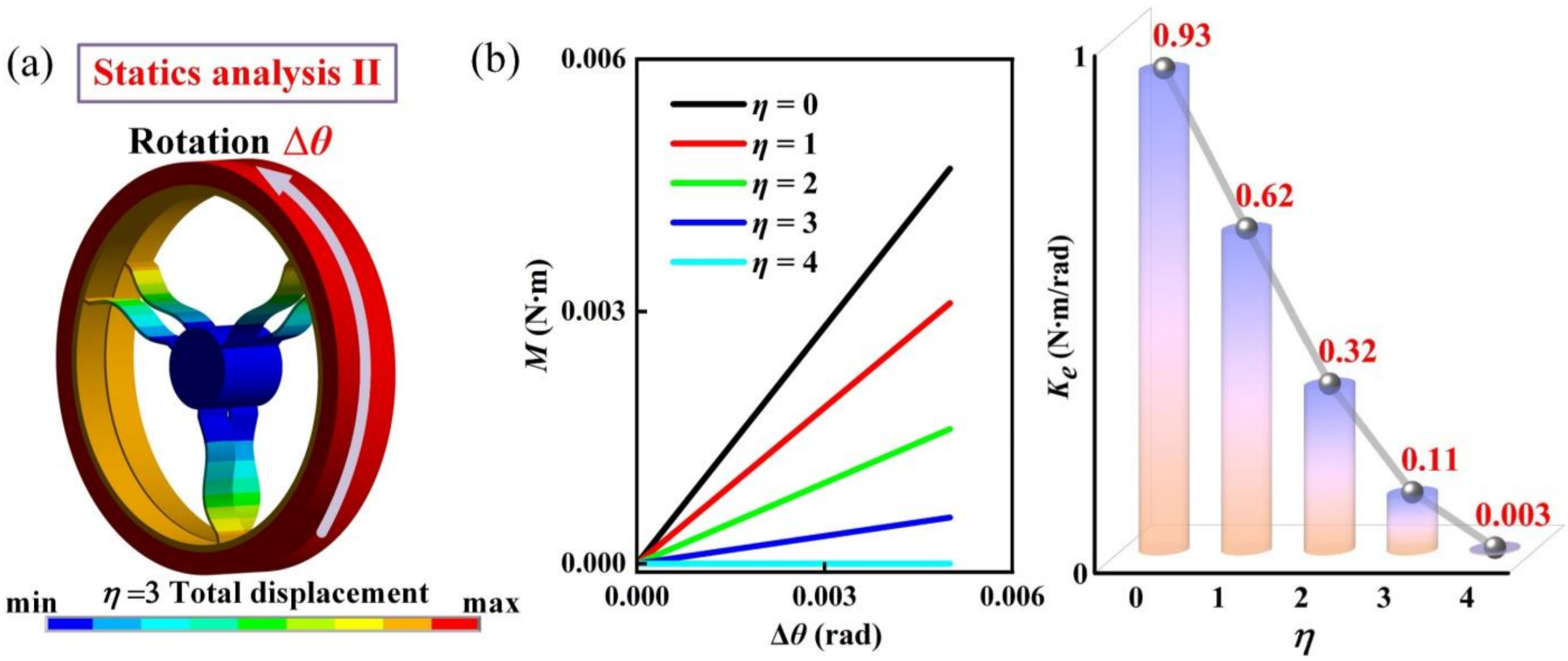

**Fig. 8.** Effective torsional stiffness of the SLG resonator for different values of $\eta$. (a) The second step static analysis by applying another rotation angles $\Delta\theta$ in the same direction to all nodes on the surfaces of both circular frames. (b) $M$–Δθ curves and the corresponding effective torsional stiffness of the SLG resonators for different values of $\eta$.

The second step of the static analysis is performed on the deformed model from the first step, as shown in Fig. 8(a). In this analysis, a small rotation angle $\Delta\theta$ is applied to replace the previous applied rotation angle on both frames, and the left and right end faces of the uniform shaft are fixed. The resulting $M$–$\Delta\theta$ curves and the corresponding effective torsional stiffness of the SLG resonators for different $\eta$ values are shown in Fig. 8(b). Two important conclusions can be drawn from these results. First, for small rotation angles, the $M$–$\Delta\theta$ curves exhibits linear relationship, where the slope of each curve corresponds to the effective torsional stiffness of the SLG resonator at the respective $\eta$. Second, the effective torsional stiffness $K_e$ of the SLG resonator decreases as the rotated teeth number $\eta$ increases, approaching zero at $\eta = 4$.

In summary, the SLG mechanism enables highly precise tuning of the effective torsional stiffness of the SLG resonator. By incrementally shifting the SLG teeth, the deformation of the curved beam can be precisely controlled, thereby enabling low-stiffness characteristics of the SLG resonator and providing an effective strategy for tailoring the dynamic response of shaft systems.

## 4. Dynamic modeling and analysis

Since the gear-tuning meta-shaft is composed of periodically arranged unit cells, the material and design parameters of each unit cell collectively determine the overall performance of the meta-shaft. This section determines the analytical dispersion relation for the meta-shaft using the transfer matrix method to define its theoretical band gap range, with particular emphasis on the tunable mechanism of the band gap characteristics due to the SLG teeth shift. Finally, FE simulations and dynamic experiments are performed to validate the robustness and tunability of the gear-tuning meta-shaft for vibration suppression.

### *4.1. Dispersion relationship and band gap analysis of the gear-tuning meta-shaft*

In periodic metastructures, the formation of torsional band gaps indicates that the propagation of torsional waves is effectively prohibited within specific frequency ranges. The identification and prediction of such band gaps provide a reliable means to estimate the attenuation regions of torsional vibrations, thereby establishing a direct link between the dispersion characteristics of the structure and its performance at suppressing vibration. Fig. 9 shows a sketch of the infinite length gear-tuning meta-shaft, which is formed by the periodic arrangement of unit cells.

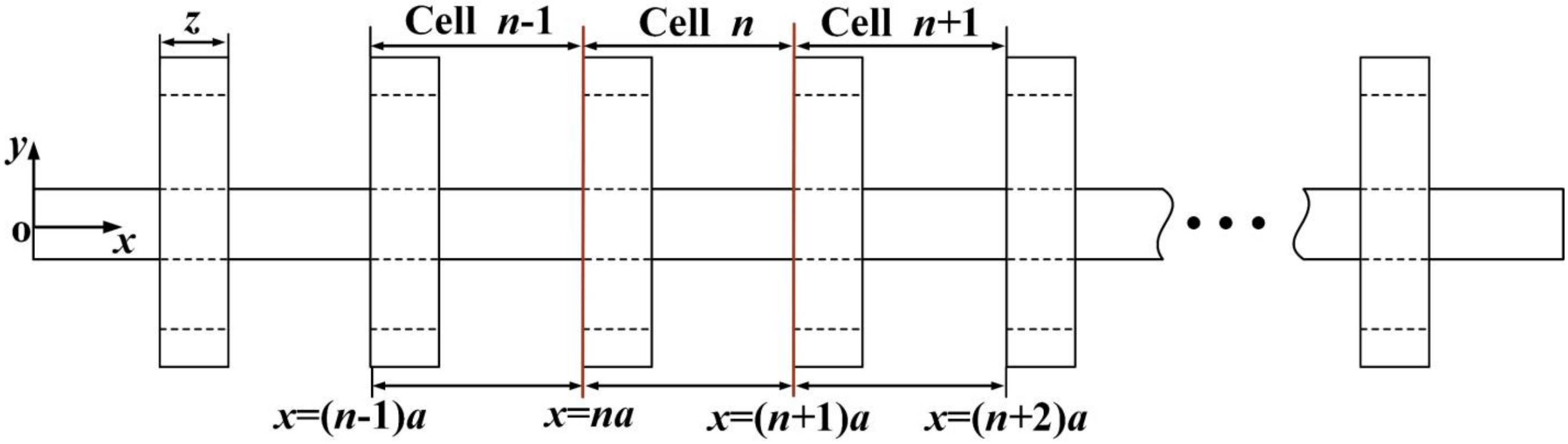

Fig. 9. Sketch of the gear-tuning meta-shaft.

The torsional vibration equations for the gear-tuning meta-shaft can be written as [27]:

$$\frac{\partial^2\theta}{\partial t^2}=c^2\frac{\partial^2\theta}{\partial x^2} \tag{21}$$

where $c$ is the wave velocity, which can be expressed as:

$$c=\sqrt{\frac{G}{\rho}} \tag{22}$$

where $G$ and $\rho$ are the shear modulus and density of the uniform shaft in the center of the gear-tuning meta-shaft, respectively.

The general solution for the $n$th unit cell can be written as:

$$\theta(x_n,t)=\left[A_n\sin(qx_n)+B_n\cos(qx_n)\right]e^{i\omega t} \tag{23}$$

where $x_n = x - na$, $\omega$ is the circular frequency, and $q$ is the wave number.

The general solution of the $n$th SLG resonator is:

$$\psi_n(t)=\Phi_n e^{i\omega t} \tag{24}$$

where $\Phi_n$ is the vibration amplitude of the $n$th SLG resonator.

The undamped motion equation of the $n$th SLG resonator can be written as:

$$J\frac{\partial^2\psi_n}{\partial t^2}+K_{\mathrm{e}}\left[\psi_n(t)-\theta\left(x_n,t\right)\right]=0 \tag{25}$$

where $J=\dfrac{\pi\rho z\left(R^2+R_1^2\right)\left(R_1^2-R^2\right)}{2}$ is the moment of inertia of the $n$th SLG resonator. $K_{\mathrm{e}}$ denotes the effective torsional stiffness of the SLG resonator, as presented in Fig. 8(b).

Substituting Eqs. (23) and (24) into Eq. (25), the vibration amplitude $\Phi_n$ of the $n$th SLG resonator can be obtained as:

$$\Phi_n=\frac{K_{\mathrm{e}}\left(A_n\sin\left(qx_n\right)+B_n\cos\left(qx_n\right)\right)}{K_{\mathrm{e}}-J\omega^2} \tag{26}$$

Applying the continuity conditions of rotation angle and torque at the interfaces between unit cell $n$–1 and $n$, we can obtain:

$$\begin{aligned}B_n&=A_{n-1}\sin(qa)+B_{n-1}\cos(qa)\\A_n&=A_{n-1}\cos(qa)-B_{n-1}\sin(qa)+\Omega B_n\end{aligned} \tag{27}$$

where $\Omega=\dfrac{K_{\mathrm{e}}J\omega^2}{GJ_cq(K_{\mathrm{e}}-J\omega^2)}$. $J_{\mathrm{c}}$ is the torsional constant of the uniform shaft, which can be written as:

$$J_{\mathrm{c}}=\frac{\pi r^2}{2} \tag{28}$$

The continuity conditions for the rotation angle and torque can be written in matrix form as:

$$\mathbf{\Psi}_n=\mathbf{T}\mathbf{\Psi}_{n-1} \tag{29}$$

where $\mathbf{T}=\begin{bmatrix}-\Omega\sin(qa)+\cos(qa) & -\Omega\cos(qa)-\sin(qa)\\ \sin(qa) & \cos(qa)\end{bmatrix}$ and $\mathbf{\Psi}_n=\left[A_n,B_n\right]^T$

For an infinite periodic gear-tuning meta-shaft, Bloch's theorem can be applied as:

$$\mathbf{\Psi}_n=e^{ika}\mathbf{\Psi}_{n-1} \tag{30}$$

where $k$ is the wave vector in the $x$-direction. Substituting Eq. (29) into Eq. (30) generates the following eigenvalue problem:

$$\left|\mathbf{T}-e^{ika}\mathbf{I}\right|=0 \tag{31}$$

where **I** is a 2 × 2 unit matrix.

For any given $\omega$, solving Eq. (31) readily yields a pair of conjugate complex solutions of the Bloch wave vector $k$, which can be used to represent the band structures [27]. Based on whether the wave vector solution has a zero or a non-zero imaginary part, it can be determined whether the torsional wave at a given frequency can propagate (band pass) within the gear-tuning meta-shaft or be restricted (band gap). Therefore, the real and complex band structures describing the relationship between the wave vector and frequency can be obtained, which respectively reflect the position and width of the band gap as well as the attenuation capability of torsional waves within the band gap. The band gap structure of the gear-tuning meta-shaft is plotted for different SLG teeth shift numbers $\eta$, as calculated using Eq. (31). The structural parameters used in the calculation are the lattice constant ($a$ = 60 mm) and the thickness of the SLG resonator ($z$= 2$b$), and the remaining material and geometric parameters are consistent with those in Section 3.

The real band structure of the gear-tuning meta-shaft based on theoretical analysis, as shown in Fig. 10(a), when the SLG is not shifted ($\eta$ = 0). The purple shaded region in the figure represents the complete torsional band gap, which spans the frequency range from 61.3 Hz to 531.5 Hz. The absolute width of this complete band gap is $\Delta f$ = 470.3 Hz, with the central frequency of the band gap at $f_c$ = 296.4 Hz. Fig. 10(b) and (c) show the real band structures of the gear-tuning meta-shaft for $\eta$ = 1 and $\eta$ = 3, respectively. It can be observed that as $\eta$ increases, both the band gap position and central frequency shift from the high-frequency region to the low-frequency region. Specifically, for $\eta$ = 1, the torsional band gap frequency range is 50.2 Hz to 428.6 Hz; for $\eta$ = 3, the torsional band gap frequency range shifts to 21.7 Hz to 182.8 Hz. This phenomenon can be attributed to the conclusion drawn in Section 3: as the value of $\eta$ increases, the effective torsional stiffness of the SLG resonator gradually decreases, which in turn shifts the torsional band gap of the gear-tuning meta-shaft toward lower frequencies.

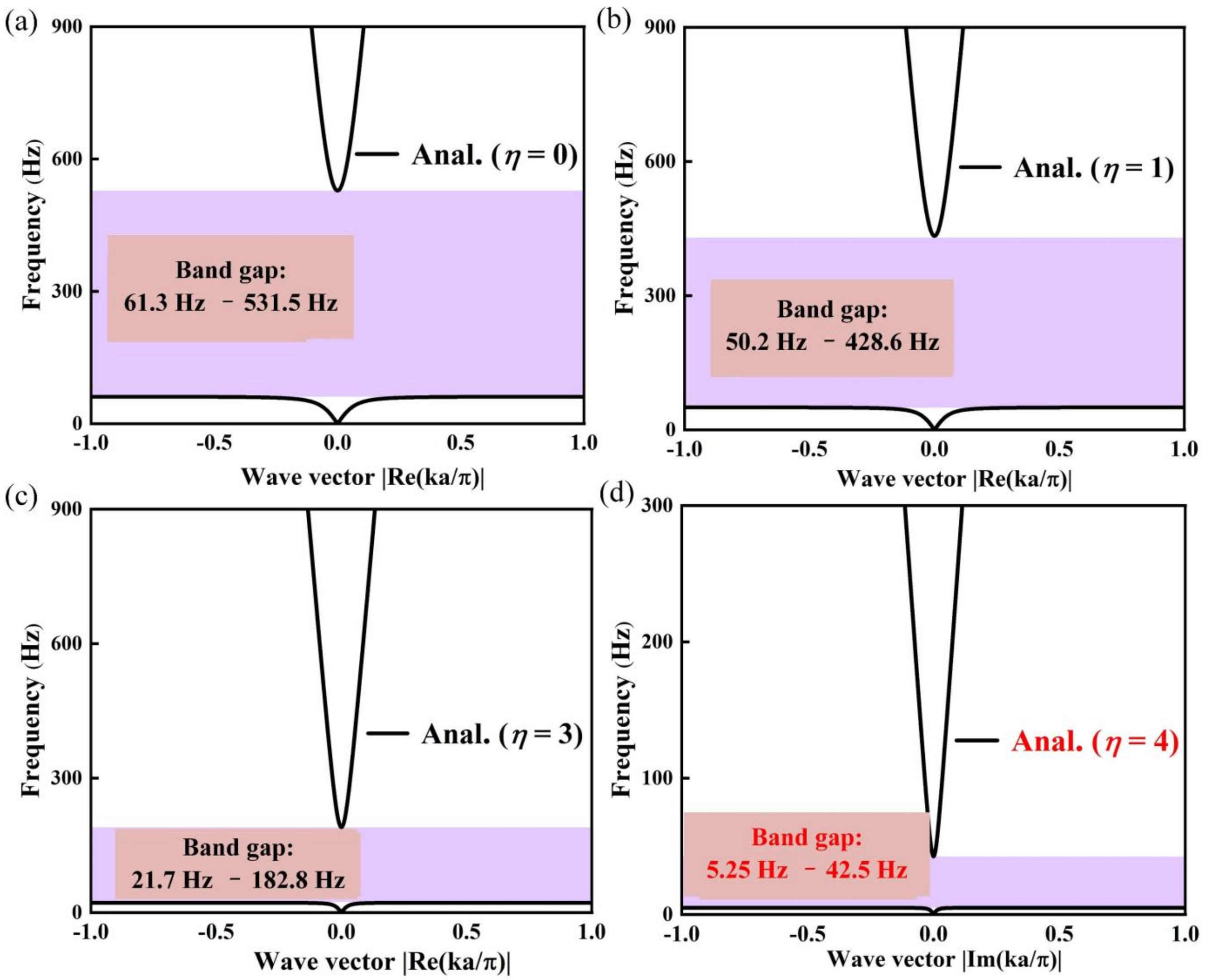

**Fig. 10.** Band gap structure of the gear-tuning meta-shaft for different values of $\eta$.(a) The real band gap structure for non-shifted SLG ($\eta$ = 0). The purple shaded region represents the complete torsional band gap. (b) Band gap structure of the SLG shifted by one tooth ($\eta$ = 1). (c) Band gap structure of the SLG shifted by three teeth ($\eta$ = 3). (d) Absolute value of the imaginary part of the wave vector.

Finally, the attenuation capability and the bandwidth of the band gap are illustrated by the complex band structure. From the imaginary part of the wave vector shown in Fig. 10(d), it can be observed that as the value of $\eta$ increases, the width of the band gap gradually narrows and the depth of the band gap decreases. This indicates that as $\eta$ increases, the attenuation ability of the band gap region in the gear-tuning meta-shaft for torsional vibrations weakens. In conclusion, the gear-tuning mechanism in the meta-shaft makes it possible to tune its torsional band gap characteristics.

### *4.2. Numerical simulation and experimental investigation of dynamics*

In this section, we further validate the theoretical band gap structure by analyzing the torsional vibration transmission characteristics, which are obtained from dynamic FE simulations and experimental testing. Fig. 11 illustrates the three main steps of the dynamic FE simulations in detail for a gear-tuning meta-shaft composed of six-unit cells. The element type and mesh size of the FE model are consistent with those described in Section 3.2. The specific steps of the dynamic analysis are as follows: (1) A nonlinear static analysis is performed by applying pre-torsion to the SLG resonators to achieve low torsional stiffness. The six resonators are subjected to the same loads and boundary conditions as in Section 3.4, with both ends of the meta-shaft possessing only one rotational degree of freedom along the $z$-axis. (2) A modal analysis is subsequently conducted on the deformed meta-shaft by performing a restart analysis, during which the boundary conditions and loads applied to all resonators are removed. (3) A harmonic response analysis is carried out using the modal superposition method on the deformed meta-shaft. During the harmonic response analysis, a harmonic excitation torque with an amplitude of 0.1 N·m about the $z$-axis is applied to the left end of the deformed meta-shaft.

The transmissibility $T$ of angular acceleration is used to evaluate torsional vibration attenuation performance of the gear-tuning meta-shaft, which can be defined as:

$$T = 20\log_{10}\left(A_{\text{out}} / A_{\text{in}}\right) \tag{32}$$

where $A_{\text{in}}$ and $A_{\text{out}}$ represent the angular acceleration amplitudes at the left and right ends of the meta-shaft, respectively.

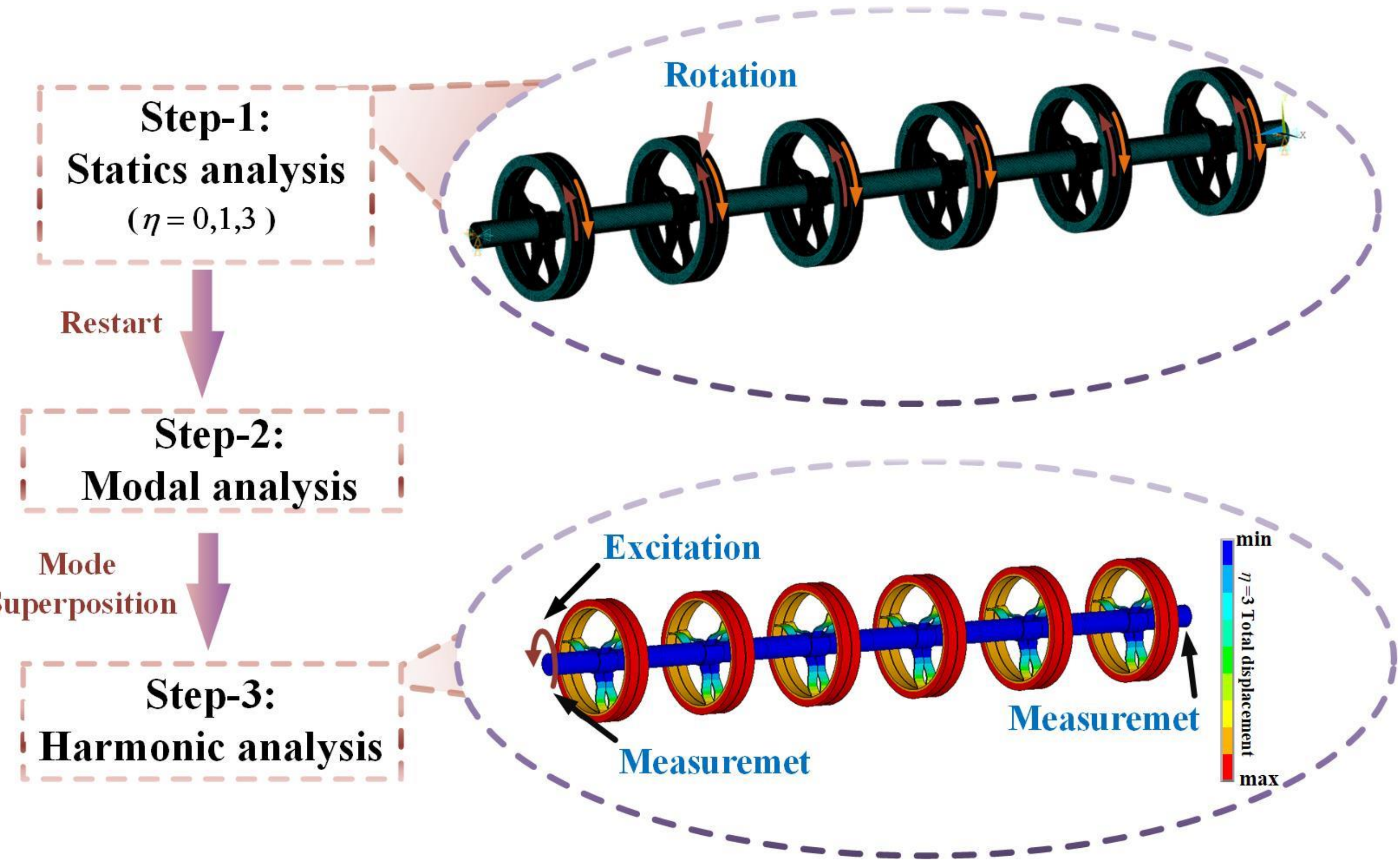

**Fig. 11.** Dynamic FE analysis flowchart of the gear-tuning meta-shaft consists of three steps: static analysis, modal analysis, and harmonic analysis.

Furthermore, a dynamic experimental platform was established, with the experimental apparatus and testing system shown in Fig. 12(a). The dynamic experiment used a rack-pinion mechanism to convert translational excitations into torsional excitations [65]. The translational excitations were generated by an electromechanical shaker (LDS V406 from Bruel & Kjaer VTS Ltd.). The pinions were fixed at both ends of the uniform shaft in the gear-tuning meta-shaft. The rack was connected to the pinion and fixed to the shaker to maintain its translational freedom. Because the pinion and rack require high precision, they were both fabricated by computerized numerical control. Additionally, the ends of the gear-tuning meta-shaft were fixed with rolling bearings on rigid supports, allowing the shaft to rotate only around its axis. As shown in Fig. 12(b), the gear-tuning meta-shaft fabricated using a 3D printer with photosensitive resin material. The geometric and material parameters were consistent with those provided in Section 3. The fixed ring of the SLG resonator was embedded with six copper nuts (as shown in Fig. 12(c)), and screws were used inside the copper nuts to fix the SLG resonators along the uniform shaft. This design facilitates the recycling of the SLG resonators in practical engineering applications.

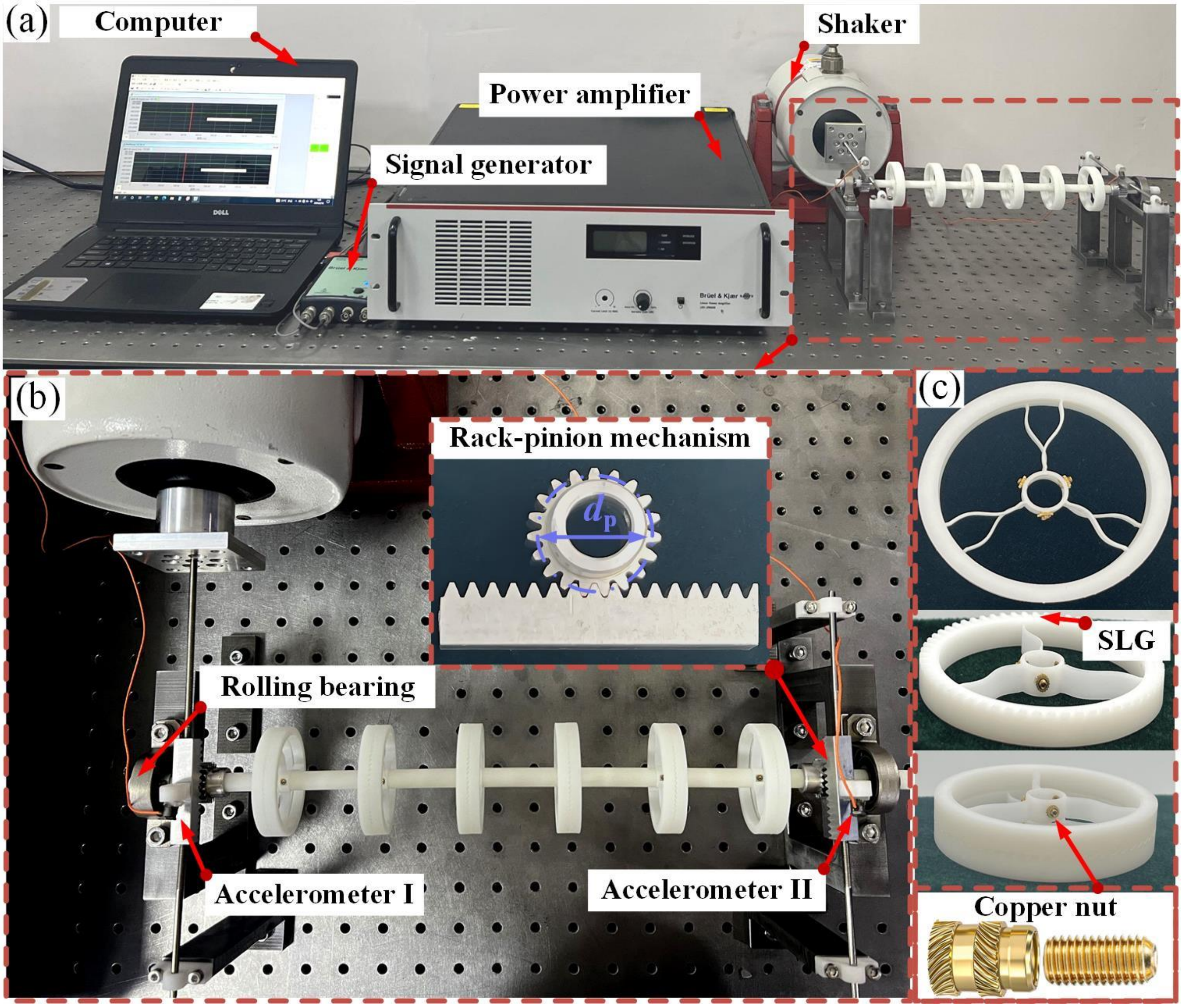

**Fig. 12.** Dynamic experimental platform of the gear-tuning meta-shaft. (a) Experimental setup for the dynamic experiments. (b) Assembly details of the gear-tuning meta-shaft, including signal generator, shaker, power amplifier, accelerometers, rolling bearing, rack-pinion mechanism, and computer. (c) A SLG resonator with embedded copper nuts, where screws secure the resonator onto the uniform shaft.

Subsequently, a white noise signal was produced by the signal generation module included in the dynamic signal collection system (PHOTON+ from Bruel & Kjaer VTS Ltd.), and then amplified by a power amplifier (LDS LPA600 from Bruel & Kjaer Sydney VTS Ltd.) to drive the electromechanical shaker. Note that the root mean square of the excitation signal should be kept at a low level to ensure that the vibration amplitude of the gear-tuning meta-shaft remains sufficiently small to satisfy the linearity assumption [66]. Two identical accelerometers (4516 from Bruel & Kjaer VTS Ltd.) were mounted on the top surface of the rack, with accelerometer I connected to the shaker to capture the input signal and accelerometer II positioned at the measurement end to record the output signal. The input and output angular acceleration amplitudes $A_{\text{in}}$ and $A_{\text{out}}$ are as follows:

$$\begin{aligned} A_{\text{in}} &= 2a_{\text{I}} / d_{\text{p}} \\ A_{\text{out}} &= 2a_{\text{II}} / d_{\text{p}} \end{aligned} \tag{33}$$

where $a_{\text{I}}$ and $a_{\text{II}}$ represent the amplitudes of tangential acceleration obtained from accelerometer I and II, respectively, and $d_{\text{p}}$ denotes the pitch circle diameter of the pinion in Fig. 12(b).

From Eqs.(32) and (33), the transmissibility of the gear-tuning meta-shaft is obtained when the SLC teeth are non-shifted ($\eta$ = 0). Then, by shifting the teeth of the six resonators on the meta-shaft prototype once ($\eta$ = 1) and three times ($\eta$ = 3), the transmissibility of the meta-shaft can be obtained for each case.

### *4.4. Dynamic results and analysis*

In this section, the results of FE simulations and experiments are compared with the theoretically predicted band gaps, further demonstrating that the gear-tuning meta-shaft allows tunability of the torsional vibration attenuation region and achieves low-frequency suppression under the gear-tuning mechanism. In Fig. 13, transmissibility curves obtained from experiments and FE simulations are represented by black solid lines and red dashed lines, respectively, while the purple shaded regions denote the torsional band gaps predicted by theoretical analysis. To facilitate direct comparison, Table 3 lists the torsional band gaps and corresponding attenuation regions derived from three results across different $\eta$ values.

The transmissibility curve of the non-shifted SLG in the gear-tuning meta-shaft is calculated based on the obtained FE simulation and experimental data, as shown in Fig. 13(a). The simulation results show an attenuation region between 60.5 and 528 Hz (center frequency: 294 Hz), while the experimental results range from 73 to 495 Hz (center frequency: 284 Hz). Due to manufacturing tolerances and gear meshing offsets in the experiment, the resonant frequencies of the each SLG resonators become non-identical. As a result, the superposition of the band gaps, together with the peak-smoothing effect introduced by damping, causes the initial decrease in the transmissibility within the attenuation region to be slower than the rapid drop observed in the FE simulation. Consequently, the onset frequency of the attenuation region in the experiment becomes slightly higher than that in the FE simulation, although the overall agreement between the two transmissibility curves remains excellent. Moreover, the attenuation regions obtained from the FE simulations and experiments are in good agreement with the torsional band gaps predicted by the theoretical analysis, with some deviations arising from additive manufacturing and assembly errors. Furthermore, Fig. 13(b) shows the total displacement fields of the gear-tuning meta-shaft at frequencies outside (point A) and inside (point B) the band gap. For point A (45 Hz), stable torsional waves propagate along the uniform shaft. However, at point B (65 Hz), most of the vibrational energy is concentrated within the SLG resonators due to their resonant response. At this starting frequency of the attenuation region in the transmissibility curve (i.e., the lower edge of the band gap), the uniform shaft and the SLG resonators oscillate in phase, which is consistent with the well-known physical mechanism of band-gap formation. Under this condition, the strong resonance of the SLG resonators suppresses the propagation of torsional waves along the uniform shaft of the gear-tuning meta-shaft.

Subsequently, the SLG teeth of the six SLG resonators are shifted individually. When the SLG shifts by one tooth ($\eta$ = 1), the simulation results show an attenuation region between 49.8 and 426.3 Hz (center frequency 238 Hz), whereas the experimental results identify an attenuation region between 65 and 405 Hz (center frequency 235 Hz), as shown in Fig. 13(c). When the SLG teeth shift three times ($\eta$ = 3), the simulation results show an attenuation region between 20.5 and 171.8 Hz (center frequency: 96.15 Hz), while the experimental results reveal an attenuation region between 35 and 164 Hz (center frequency: 99.5 Hz), as shown in Fig. 13(d). These results indicate that as $\eta$ increases, both the initial and central frequencies of the attenuation region shift from the high-frequency to the low-frequency range, thereby enabling the suppression of low-frequency torsional vibrations. Specifically, when shifting by one and three teeth, the central frequency of the torsional vibration attenuation region is reduced by 17.25% and 64.96%, respectively, compared to the non-shifted case. Moreover, with increasing $\eta$, the attenuation ability of the band gap region in the gear-tuning meta-shaft for torsional vibrations weakens, which is consistent with the conclusion drawn from the complex band gap structure in section 4.1.

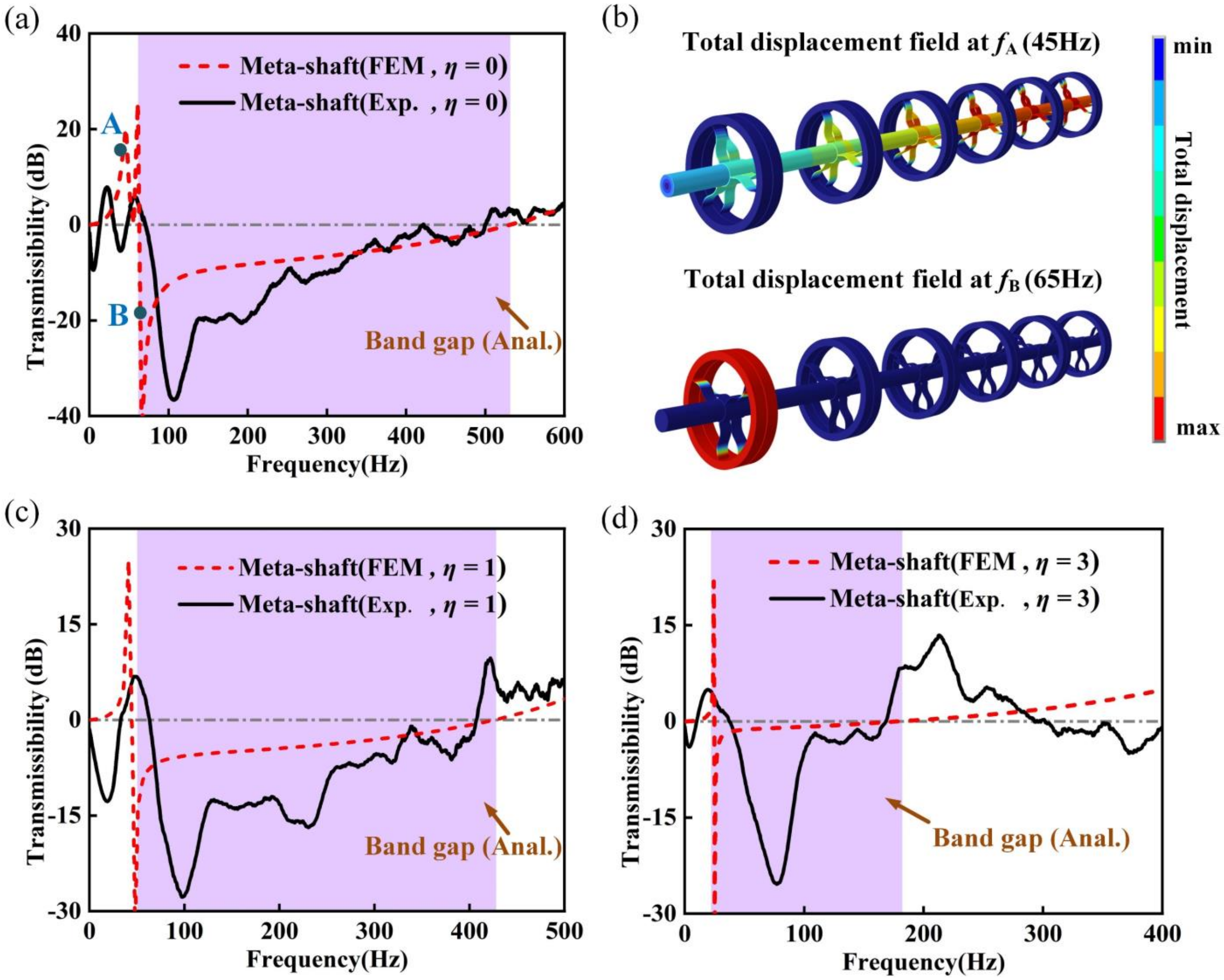

**Fig. 13.** Transmissibility curves obtained from experiments and FE simulations under different $\eta$. (a) Comparison of FE simulations (marked in red dashed lines) and dynamic experiments (marked in black solid lines) for $\eta$ = 0, with purple shading indicating theoretical band gaps. The attenuation regions obtained from the FE simulation (60.5 ~ 525.4 Hz) and the experiment (73 ~ 495 Hz). (b) Total displacement field of the gear-tuning meta-shaft ($\eta$ = 0) at 45 Hz and 65 Hz. (c) Transmissibility curves for $\eta$=1: FE simulation (49.8 ~ 426.3 Hz) and experiment (65 ~ 405 Hz). (d) Transmissibility curves for $\eta$=2: FE simulation (20.5 ~ 171.8 Hz) and experiment (35 ~ 164 Hz).

**Table 3**

Torsional band gaps and attenuation regions for different $\eta$.

| Shift count of the SLG ($\eta$) | $\eta = 0$ | $\eta = 1$ | $\eta = 3$ |
|---|---|---|---|
| Band gap (Analytical) | 61.3 ~ 531.5 Hz | 50.2 ~ 428.6 Hz | 21.7 ~ 182.8 Hz |
| Attenuation region (FEM) | 60.5 ~ 525.4 Hz | 49.8 ~ 426.3 Hz | 20.5 ~ 171.8 Hz |
| Attenuation region (Experiment) | 73 ~ 495 Hz | 65 ~ 405 Hz | 35 ~ 164 Hz |
| Center frequency of the attenuation region (FEM) | 293 Hz | 238 Hz | 96.2 Hz |
| Center frequency of the attenuation region (Experiment) | 284 Hz | 235 Hz | 99.5 Hz |
| Relative error in the center frequency of the attenuation region. (Experiment & FEM) | 3.1 % | 1.3 % | 3.4 % |
| Relative decrease in the center frequency of the attenuation region (Experiment) | ~ | 17.25 % | 64.96 % |

Overall, the consistency observed among theoretical analysis, numerical simulations, and experiments demonstrates the effectiveness and robustness of the proposed tunable meta-shaft design concept. This design enables precise control over the attenuation region of torsional vibrations/waves by shifting the SLG teeth, making it a versatile tool for adjusting the dynamic response of the shaft system in engineering applications such as marine propulsion systems, wind-turbine drive trains and automotive drivetrains, where the suppression of low-frequency torsional vibrations is of critical importance. The ability to tune the attenuation region provides crucial theoretical insights into the effective suppression of low-frequency torsional vibrations and their adaptation to various operating conditions.

## 5. Conclusion

In this study, we propose a gear-tuning meta-shaft with self-locking gear (SLG) resonators, where a simple gear-tuning mechanism replaces complex tuning methods to enable precise regulation of the vibration attenuation frequency and to realize robust suppression of low-frequency torsional vibrations. As the key components of the SLG resonator, six inner curved beams with sine profiles incorporating an initial inclination angle are designed to provide tunable torsional stiffness for the resonator through their deformed shapes controlled by shifting the SLG teeth on the edge of the resonator. A static model for calculating the torsional stiffness of the curved beam-shaft unit is developed using the CBCM method, and its validity is verified through numerical simulations and experiments. Meanwhile, it is demonstrated that by adjusting the initial inclination angle and apex height of the sine curve, the mechanical properties of the curved beam-shaft unit can be tuned, thereby enabling the realization of positive, near-zero, and negative torsional stiffness.

In addition, based on the gear-tuning mechanism, the SLG resonator can achieve resonant frequency modulation and open tunable low-frequency torsional band gaps consistent with theoretical predictions. Subsequently, a prototype of the gear-tuning meta-shaft is fabricated by attaching SLG resonators to a uniform shaft, and its transmissibility was measured after each shift of the SLG teeth on the edge of the resonator. The dynamic experimental and numerical simulation results indicate that the attenuation ranges of torsional vibrations are consistent with the predicted band gaps. Specifically, when SLG teeth are shifted by one tooth and three teeth, the center frequency of the torsional vibration attenuation region is reduced by 17.25% and 64.96%, respectively, compared to the non-shifted case.

Achieving both broadband and low-frequency performance in metastructures has long been a challenge. Low-frequency tunability is achieved through a simple gear-tuning mechanism, while this approach also offers potential for broadband performance. In our future work, we will continue exploring tunable and low-frequency torsional vibration suppression, aiming to integrate it with innovative smart materials/structures to advance research on intelligent tunable wide and low-frequency band gaps.

## Acknowledgments

This work was supported in part by the National Natural Science Foundation of China (NSFC) under Grants No. U22B2078. Y.C. acknowledges the support from the Hong Kong Research Grants Council (Grant No. RGC-GRF 16204824).